\documentclass[prl, aps, twocolumn, floatfix, superscriptaddress, amsmath,  longbibliography]{revtex4-2}
\usepackage{graphicx}
\usepackage{graphics}
\usepackage{mathtools}
\usepackage{float}
\usepackage{amssymb}
\usepackage{bbold}

\usepackage[dvipsnames]{xcolor}
\definecolor{darkblue}{rgb}{0.0, 0.0, 0.75}

\usepackage{color}
\usepackage[colorlinks=true,
            linkcolor=darkblue,
            urlcolor=darkblue,
            citecolor=darkblue]{hyperref}

	\usepackage[normalem]{ulem} 
	\definecolor{mgreen}{RGB}{1,123,0}

\def \br{{\bf r}}

\def \mJ{\text{J}}
\def \ms{\text{s}}

\def \md{\mathrm{d}}
\def \mI{\mathrm{I}}

\def \mB{\mathrm{B}}
\def \ms{\mathrm{s}}
\def \mm{\mathrm{m}}

\def \mum{\mu\mathrm{m} }

\def \cD{\mathcal{D}}

\def \mpump{\mathrm{pump}}
\def \msignal{\mathrm{signal}}

\def \mHz{\mathrm{Hz}}

\def \mms{\mathrm{ms}}

\def \cD{{\cal{D}}}

\def \tI{\tilde{I} }

\begin{document}
\title{Atomic Josephson Parametric Amplifier}
\author{Vijay Pal Singh}
\affiliation{Quantum Research Centre, Technology Innovation Institute, Abu Dhabi, UAE}
\author{Luigi Amico}
\affiliation{Quantum Research Centre, Technology Innovation Institute, Abu Dhabi, UAE}
\affiliation{Dipartimento di Fisica e Astronomia, Universit\`a di Catania, Via S. Sofia 64, 95123 Catania, Italy}
\affiliation{INFN-Sezione di Catania, Via S. Sofia 64, 95127 Catania, Italy}
\author{Ludwig Mathey}
\affiliation{Zentrum f\"ur Optische Quantentechnologien and Institut f\"ur  Quantenphysik, Universit\"at Hamburg, 22761 Hamburg, Germany}
\affiliation{The Hamburg Centre for Ultrafast Imaging, Luruper Chaussee 149, Hamburg 22761, Germany}
\date{\today}
%

%
\begin{abstract}
We study the dynamics of a driven atomic Josephson junction that we propose as a parametric amplifier.
By periodically modulating the position of the barrier, we induce a small current across the junction, serving as our input signal. 
The pump field is implemented by modulating  the barrier height at twice the Josephson plasma frequency. 
The resulting dynamics exhibit parametric amplification of the signal through nonlinear mixing between the signal and pump fields, which is encoded in a  specific microscopic pattern of density waves and phase excitations that can be addressed within the  experimental cold atoms capabilities. 
This work paves the way for tunable amplifiers in atomtronic circuits, 
with potential applications in several fields including precision measurements and quantum information processing. At the same time, our analysis provides  the microscopic explanation of the general notion of parametric amplification occurring in nonlinear coherent devices.
\end{abstract}
\maketitle
%
%

Parametric amplification is a fundamental concept in nonlinear physics: When the  system's  natural frequency $\omega_0$ is periodically modulated in time with frequency $\omega_\md$ (the pump), a nearly resonant condition can be achieved for  $\omega_\md \approx 2 \omega_0$; 
in this case, an energy transfer from the pump to the system is induced \cite{Case1996}. 
This idea was implemented for classical electronics \cite{Roer1994} and then extended to 
quantum optics \cite{Mollow1967, Baumgartner1979}, semiconductors \cite{Thomson1976} and superconducting materials \cite{Yurke87,Yurke1989, Siddiqi2004, Tholen2007, Yamamoto2008}. Parametric amplifiers are extensively used in important technological sectors as  radio astronomy \cite{asaki2023astronomical}, communications \cite{marhic2015fiber}, radar systems \cite{dvornikovamplifiers}, spectroscopy \cite{koch2023terahertz} and sensing \cite{Du2018}.
In quantum technology, parametric amplifiers have become relevant for many applications ranging from the generation of squeezed states \cite{Yurke87,Yurke1989, Zagoskin2008}, entanglement and nonclassical states \cite{Ou1990, Nagasako2001}, and quantum-limited amplification \cite{Castellanos-Beltran2008} to quantum computing \cite{Blais2021}, quantum sensing and metrology \cite{degen2017quantum,bass2024quantum}.

Superconducting  Josephson junctions provide non linear circuits that have been thoroughly employed to achieve a parametric amplification.  
These devices amplify weak microwave signals with minimal added noise thus generating high-quality amplification.  
When we apply a strong pump field with frequency $\omega_\md$ at twice the Josephson plasma frequency
$\omega_0$, i.e., $I_{\mpump}(t) \sim \sin(2\omega_0)$, and a weak signal field $I_{\msignal}(t) \sim \cos(\omega_\ms t)$, 
with frequency $\omega_\ms \approx \omega_0$,  the nonlinear inductance of the Josephson junction produces parametric amplification
through a nonlinear mixing process between the signal and pump fields \cite{Macklin2015}. 
$\omega_0$ is determined by $\omega_0= \sqrt{2e I_c/(\hbar C)}$, where $2e$ is the electronic charge, $I_c$ the critical current, 
and $C$ the capacitance. 
The idler modes at frequency $\omega_{\mI} =  \omega_\md- \omega_s$ ensure the energy conservation. 
The signal mode at $\omega_\ms$ is amplified as the pump energy is transferred to it through the nonlinear mixing process.

Ultracold atoms have emerged as an ideal platform for implementing and exploring atomtronic analogs of superconducting circuits 
 \cite{Amico2005, Amico2021, Amico2022, Ramanathan2011, Eckel2014, Ryu2013,  Ryu2020, Chien2015, Krinner2017, Polo2024}. 
One breakthrough was the realization of Josephson junctions (JJs) using weak links of ultracold atom clouds  \cite{Cataliotti2001,Albiez2005, LeBlanc2011, Spagnolli2017,  Pigneur2018}. 
This has not only enabled the study of important quantum effects, such as dc-ac Josephson effects \cite{Levy2007, Kwon2020, Pace2021}, 
current-phase relation  \cite{Luick2020, Kwon2020}, Josephson-tunneling enhanced superfluidity \cite{Rydow2024}, 
with remarkable accuracy and control of the physical conditions, 
but has also laid the foundation for the development of atom-based devices, such as atomic quantum interference devices (AQUIDs) 
\cite{Wright2013, Eckel2014,Ryu2013, Campbell2014, Ryu2020}.  
Theoretical studies have examined Josephson effects \cite{Smerzi1997,Raghavan1999,Giovanazzi2000}, 
the implementation of universal set of logic gates \cite{Jahrling2024}, atomic conductivity in pump-probe setting \cite{Zhu2021}, or driven dynamics  \cite{Raghavan1999, Kohler2003, Eckardt2005, Grond2011}. 
Recently, quantized plateaus of chemical potential were predicted in driven atomic JJs \cite{SinghShapiro}, 
which are the analogue of Shapiro steps in superconducting JJs \cite{Shapiro1963, Grimes1968}. 
Atomic Shapiro steps were measured in Refs. \cite{Del_Pace2024,Bernhart2024}, serving as a basis for a pressure standard in atomic condensates.

\begin{figure}[t]
\includegraphics[width=1.0\linewidth]{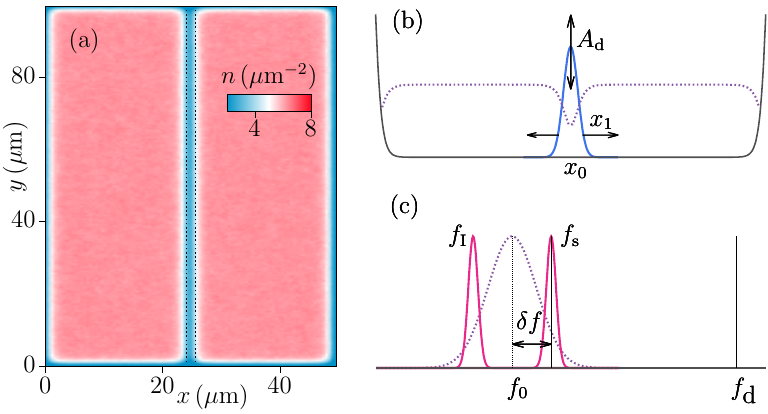}
\caption{Atomic Josephson parametric amplifier (JPA). 
(a) Simulation of a Josephson junction, which consists of two 2D clouds separated by a Gaussian tunnel barrier of height $V_0$ and width $w$ 
(indicated by two dotted vertical lines). 
We use $V_0/\mu=1.5$ and $w/\xi=1.1 $, where $\mu$ is the mean-field energy and $\xi$ is the healing length. 
(b) Sketch of the JPA protocol. 
We periodically modulate the barrier position using $x(t) = x_0 + x_1 \sin(2\pi f_\ms t)$, where $x_1$ is the amplitude and $f_\ms$ is the frequency. 
For the pump we modulate the barrier height, i.e., $V(t)= V_0\bigl(1+ A_\md \cos(2\pi f_\md t) \bigr) $, 
where $A_\md$ is the amplitude and $f_\md$ is the frequency.   
(c) Sketch of the operation of a JPA. We modulate the barrier height near twice the plasma frequency $f_0$ of the JJ. 
The signal mode at $f_\md/2 + \delta f$ is amplified and an idler mode at frequency $f_\md/2 - \delta f$ is created, 
with $ \delta f$ being the tuning parameter. 
}
\label{Fig:system}
\end{figure}

With remarkable progress in the field of ultracold atoms, driven matter-wave circuits are now well within experimental reach. 
Here, we propose the implementation of a Josephson parametric amplifier (JPA) using ultracold atoms. 
Our protocol relies on an atomic JJ in which we periodically modulate both the position and height of the tunnel barrier, see Figs. \ref{Fig:system}(a, b).
This can be realized in ultracold atom experiments using digital micromirror devices  \cite{Del_Pace2024,Bernhart2024}.  
In our system, the modulation of the barrier position serves as the input current signal, 
while the modulation of the barrier height acts as a pump field. 
By tuning the pump frequency to twice the Josephson plasma frequency, 
we induce nonlinear mixing between the pump and signal modes, 
see Fig. \ref{Fig:system}(c). A strong pump field drives the system into a regime
where parametric amplification occurs. 
Our results are obtained using classical-field dynamics, which include fluctuating bosonic fields beyond the mean-field description \cite{Blakie2008, Polkovnikov2010,  Singh2016, Singh2020sound}. 
Additionally, the system dynamics are well captured by a driven circuit model.

\begin{figure}[t]
\includegraphics[width=1.0\linewidth]{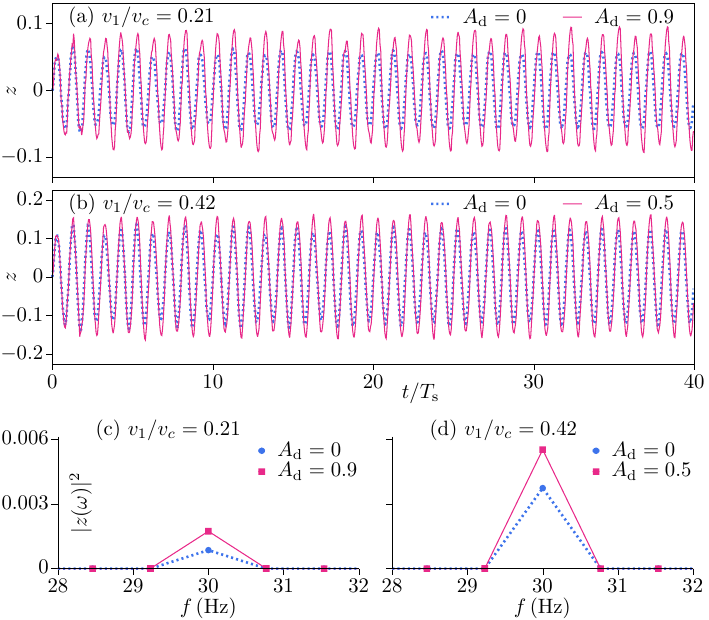}
\caption{JPA dynamics. 
Time evolution of the imbalance $z(t)$ demonstrating the amplification for the signal 
at frequency $f_\ms=30\, \mHz$ and two different sets of signal and pump amplitudes:  
(a) $v_1/v_c=0.21$ and $A_\md=0.9$. 
(b) $v_1/v_c=0.42$ and $A_\md=0.5$. 
$v_c$ is the critical velocity above which the junction is resistive. 
The results at $A_\md=0$ represent the system without the pump. 
$T_\ms =1/f_\ms$ is the signal oscillation period.  
(c, d) Power spectrum of the time evolution in (a, b). 
}
\label{Fig:dynamics}
\end{figure}

\begin{figure}
\includegraphics[width=1.0\linewidth]{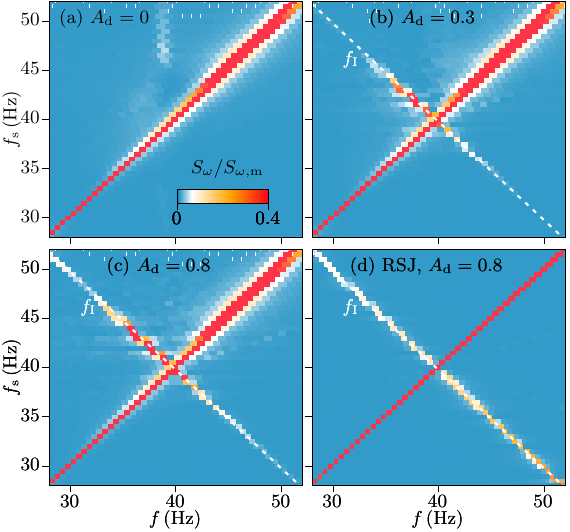}
\caption{Nonlinear mixing in the power spectrum  $S_\omega=|z(\omega)|^2$ for signal frequencies between $f_\ms=28$ and $52\, \mHz$ at  $\tI_1 \equiv I_1/I_c=0.42$. 
For each column the spectrum is normalized by its maximum value  $S_{\omega, \mm}$. 
(a) Without pump the spectrum mainly displays the central peak at $f_\ms=f$.  
(b, c) In the presence of pump there is another peak at $f_\md- f_\ms$ (indicated by white dashed line), 
located at the frequency of the signal mirrored at the plasma frequency $f_0$. 
The case of $f_\ms=f_0$, where signal and idler are at the same frequency, is a special case known as the degenerate case. 
(d) The result of the RSJ circuit model for $\tI_1=0.42$ and $A_\md=0.8$; see text.
}
\label{Fig:spectrum}
\end{figure}

{\it{System and protocol.}---} 
We consider a homogeneous cloud of bosons confined in a box of dimensions $L_x \times L_y$. 
We study the dynamics of this system using classical-field dynamics within the truncated Wigner approximation \cite{Blakie2008, Polkovnikov2010, Singh2016, Singh2020sound}.  
The system is described by the Hamiltonian
\begin{equation} \label{eq_hamil}
\hat{H}_{0} = \int \mathrm{d}{\bf r} \Big[ \frac{\hbar^2}{2m} \nabla \hat{\psi}^\dagger({\bf r}) \cdot \nabla \hat{\psi}({\bf r})  + \frac{g}{2} \hat{\psi}^\dagger({\bf r})\hat{\psi}^\dagger({\bf r})\hat{\psi}({\bf r})\hat{\psi}({\bf r})\Big].
\end{equation}
$\hat{\psi}$ ($\hat{\psi}^\dagger$) is the bosonic annihilation (creation) operator. 
The interaction $g=\tilde{g} \hbar^2/m$ is given in terms of the dimensionless parameter $\tilde{g} =  \sqrt{8 \pi} a_s/\ell_z$, 
where $m$ is the mass, $a_s$ is the s-wave scattering length and $\ell_z= \sqrt{\hbar/(m \omega_z)}$ is the harmonic oscillator length in the transverse direction. 
In the limit of a large atom number $N$, we employ the classical-field representation, 
where the operators $\hat{\psi}$ in Eq. (\ref{eq_hamil}) and the equations of motion are replaced by complex numbers $\psi$. 
The initial states $\psi (\br,t=0)$ are generated in a grand canonical ensemble with chemical potential $\mu$ and temperature $T$ via a classical Metropolis 
algorithm.
This distribution captures fluctuations of $\psi (\br,t=0)$ around its mean field value. 
Finally, each initial state is evolved according to the classical equations of motion
\begin{align}\label{eq:eom}
 i \hbar \dot{\psi}(\br, t) = \Bigl(  - \frac{\hbar^2}{2m} \nabla^2 + V(\br, t) + g|\psi|^2 \Bigr) \psi(\br, t),
\end{align}
which include the barrier potential given by $V({\bf r},t)  = V_0 (t) \exp \bigl[- 2 \bigl( x-x(t) \bigr) ^2/w^2 \bigr]$. $V_0(t)$, $w$  and $x(t)$ are the barrier's  strength, width and location.  
We perform numerical simulations by discretizing space on a lattice of size $N_x \times N_y = 100 \times 200$ with the discretization length $l=0.5\, \mu \mathrm{m}$.   
To illustrate the key idea, we focus on a concrete realization using $^{6}\mathrm{Li}_2$ molecules, 
though our protocol is applicable to any cold-atom degenerate gas. 
We set the density to $n \approx 5.6\, \mu \mm^{-2}$, interaction parameter $\tilde{g}=0.1$, and temperature ratio $T/T_0=0.06$, with system dimensions $L_x \times L_y = 50 \times 100\, \mum^2$.  
The critical temperature $T_0$ is estimated by $T_0 = 2\pi n \hbar^2/(m k_\mB \cD_c)$, 
where $\cD_c= \ln(380/\tilde{g})$ is the critical phase-space density \cite{Prokofev2001, Prokofev2002}. 
We use $w/\xi = 1.1$ and  $ V_0/\mu \equiv \tilde{V}_0 = 1.5$, 
where $\xi= \hbar/\sqrt{2m gn}$ is the healing length and $\mu = gn$ is the mean-field energy. 
To create the weak link at $x(t)= x_0= L_x/2$, we linearly ramp up $V_0$ over $200\, \mms$ and then wait for $50\, \mms$. 
The density distribution $n(x, y)=|\psi(\br, t)|^2$ is shown in Fig. \ref{Fig:system}(a), 
which is averaged over the initial ensemble.  
For the signal we periodically modulate the barrier position using $x(t) = x_0 + x_1 \sin(2\pi f_\ms t)$, 
where $x_1$ is the amplitude and $f_\ms $ is the frequency, see Fig. \ref{Fig:system}(b). 
The resulting barrier velocity $v(t)= v_1 \cos(2\pi f_\ms t)$, with $v_1= 2\pi f_\ms x_1$, induces an oscillating current across the junction.
We calculate the atom number $N_L(t)$ ($N_R(t)$) in the left (right) reservoir to determine the imbalance $z(t)= (N_L(t) - N_R(t))/N$, 
where $N=(N_L + N_R)$ is the total atom number.   
For the spectrum analysis we calculate the power spectrum 
\begin{align}
S_\omega=|z(\omega)|^2,
\end{align}
where $z(\omega) = (1/\sqrt{T_n}) \int dt  \, z(t) \exp(-i\omega t)$ is the Fourier transform of $z(t)$. $T_n$ is the sampling time for the numerical Fourier transform.
The junction displays a well-defined resonance at the Josephson plasma frequency $f_0$, 
which we determine by analyzing  $S_\omega$ at varying frequency, 
obtaining $f_0=39.4\, \mHz$ in the limit of $v_1 \rightarrow 0$  (see End Matter for details). 
For the parametric pump we modulate the barrier height, i.e.,  $V(t)= V_0\bigl(1+ A_\md \cos(2\pi f_\md t) \bigr) $, 
where $A_\md$ is the amplitude and $f_\md$ is the frequency, as depicted in Fig. \ref{Fig:system}(b).
This alters the nonlinear inductance of the junction, which gives rise to amplification of the signal due to energy exchange between different modes. 
When $V(t)$ is modulated at twice the plasma frequency, the system undergoes parametric amplification, 
where a signal at frequency $f_\md/2+\delta f$ is amplified and an idler at frequency $f_\md/2-\delta f$ is created, 
with $\delta f$ being the tuning parameter, see Fig. \ref{Fig:system}(c). 
In Fig. \ref{Fig:dynamics}, we demonstrate the parametric  enhancement for the signal at $f_\ms=30\, \mHz$ with respect to the system without pump. 
The system quickly reaches the steady state after $2-3$ signal cycles and the enhancement persists over long times as required for robust amplification. 
The power spectrum exemplifies the amplification in frequency space.


%
\begin{figure}
\includegraphics[width=1.0\linewidth]{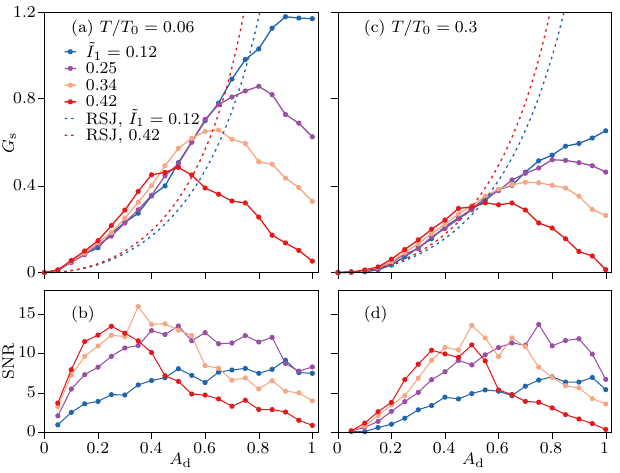}
\caption{JPA characteristics. 
(a) Dimensionless  gain $G_\ms=S_\omega(A_\md)/S_\omega(0) - 1$ as a function of $A_\md$ 
is shown for signal amplitudes between $\tI_1=0.12$  and $0.42$.
The results of the RSJ model are indicated by the dashed lines.
(b) Signal-to-noise ratio (SNR) corresponding to the results shown in (a).
(c, d)  Simulation and RSJ model results at temperature $T/T_0=0.3$,
using the same relative signal frequency with respect to the Josephson plasma frequency as in panels (a, b); see text.
%
}
\label{Fig:gain}
\end{figure}
{\it{Characterization of the JPA dynamics.}---} 
To characterize the dynamical regimes we map out the imbalance spectrum for signal frequencies between $f_\ms=20$ and $100\, \mHz$, 
which exhibits JPA resonances and other excitation peaks 
\footnote{See Supplemental Material for the driven junction spectrum, the current-chemical potential characteristic, and the driven RSJ circuit model, 
which includes Refs. \cite{SinghShapiro, Singh2020jj}.}. 
Fig. \ref{Fig:spectrum} highlights the interplay between the signal and idler modes for the lowest JPA resonance. 
The mode weights are controlled by the pump amplitude.  
To describe the JPA dynamics  we introduce a driven RSJ (resistively shunted junction) circuit model. 
The barrier velocity results in a modulation of the current across the junction given by $I_1 \cos(\omega_s t)$, 
where $I_1 = v_1 I_c/v_c$ in which $v_c$ is the critical velocity associated with the critical current $I_c$ \cite{Note1}.
The barrier height modulation changes the tunneling energy, which can be incorporated 
as $I_c\bigl( 1 + A_\md \sin(\omega_\md t) \bigr)$. 
The Kirchhoff law of the driven RSJ circuit  reads
\begin{align}
 I_1 \cos(\omega_s t) = I_c\bigl( 1 + A_\md \sin(\omega_\md t) \bigr)  \sin \phi - G \Delta \mu,  \label{eq:rsj} 
\end{align}
where $\phi= \phi_L - \phi_R$  is the phase difference across the junction and $G$ is the conductance \cite{Note1}.
The Josephson relation for the phase dynamics is $\hbar \dot{\phi} = - \Delta \mu$,  
with $\Delta \mu$ playing the role of the voltage  across the junction.
We numerically solve Eq. \ref{eq:rsj} for various signal and pump parameters and obtain the time evolution $\Delta \mu (t)$. 
The values of $I_c= 135\, (10^3/\ms)$ and $\hbar G= 59.8$ are determined from the simulated current-chemical potential relation \cite{Note1}. 
With $I_c$ and the effective charging energy $E_C = 4 (\partial \mu/\partial N)$ 
we estimate the Josephson frequency $\omega_\mJ = \sqrt{I_c E_C/\hbar} $ and 
obtain $f_\mJ = 38\, \mHz$ in  agreement with the value of $f_0$.
The result of the spectrum $|\Delta \mu(\omega)|^2$ is shown in Fig. \ref{Fig:spectrum}d, 
which captures the dynamical response of the JPA. 
We note that the circuit model also describes higher JPA resonance and other excitation peaks well \cite{Note1}.


%
\begin{figure}[t!]
\includegraphics[width=1.0\linewidth]{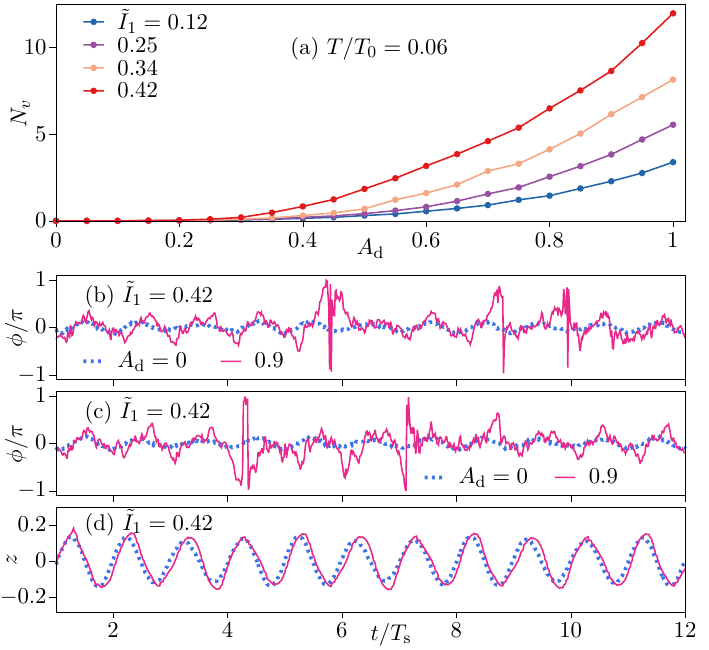}
\caption{Dephasing dynamics. 
(a) Average vortex number $N_v$ as a function of $A_\md$ for the parameters used in Fig. \ref{Fig:gain}(a). 
(b, c)  Time evolution of the junction phase $\phi(t)$ for two different samples at $\tI_1=0.42$ with (continuous line) and without pump (dashed line).
(d) Imbalance $z(t)$, averaged over many samples, corresponding to the parameters in (b, c). 
%
}
\label{Fig:dephasing}
\end{figure}
%


To quantify the overall amplification, we define a dimensionless gain $G_\ms=S_\omega(A_\md)/S_\omega(0) -1$, 
where $S_\omega(A_\md)$  represents the signal peak value and $S_\omega(0)$ corresponds to the result without pump. 
In Fig. \ref{Fig:gain}(a) we demonstrate that $G_\ms$ initially increases approximately quadratically with $A_\md$, and then reaches a maximum for a value of $A_\md$, that decreases with increasing $\tI_1=I_1/I_c$. 
For larger values of $A_\md$, the gain $G_\ms$ decreases due to the dephasing dynamics as explained in the subsequent discussion. 
While the nonlinear  increase is captured by the circuit model, 
the model fails to describe the decrease beyond the maximum, primarily visible for larger $\tI_1$. 
Fig. \ref{Fig:gain}(b) presents the signal-to-noise ratio (SNR), defined as $\bar{S}_\omega/\Delta \bar{S}_\omega$,  
where we compute the mean signal value $\bar{S}_\omega = \left\langle S_\omega(A_\md) - S_\omega(0) \right\rangle$ and the noise $\Delta \bar{S}_\omega$ from the variance determined by $\Delta \bar{S}_\omega^2  = \left\langle ( S_\omega(A_\md) - S_\omega(0) -\bar{S}_\omega )^2 \right\rangle $. Here, $\langle .. \rangle$ denotes a statistical average over $100$ samples of the ensemble.
The SNR depends on both $A_\md$ and $\tI_1$, with a visible increase with increasing $A_\md$.
The maximal regime of the SNR is in the same regime as the maximal gain $G_\ms$, 
suggesting that this is the optimal regime of operation.  
In Fig. \ref{Fig:gain}(c, d), the simulation results obtained at a higher temperature of $T/T_0=0.3$ show a noticeable reduction in overall gain, 
while still exhibiting a qualitatively similar trend to that observed at $T/T_0=0.06$.  
Thermal fluctuations suppress the Josephson plasma frequency $f_0$ and the critical current $I_c$, 
resulting in $f_0 = 35.2\, \mHz$ and $I_c= 103\, (10^3/\ms)$ for $T/T_0=0.3$, 
see End Matter and \cite{Note1} for details.

\begin{figure}[t!]
\includegraphics[width=1.0\linewidth]{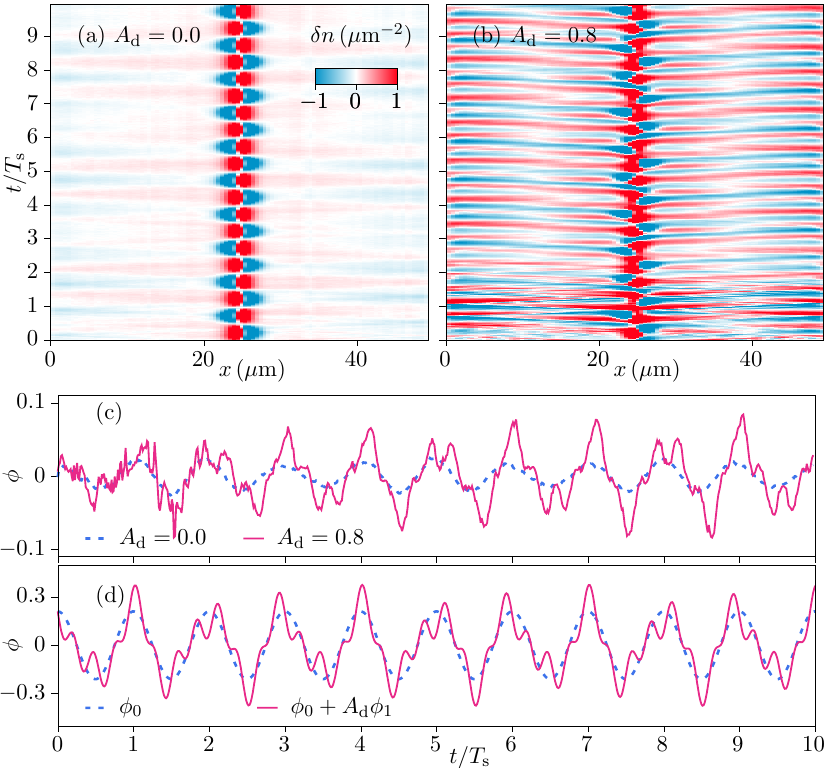}
\caption{Underlying nonlinear mechanism. 
(a) Time evolution of the density $\delta n(x, t) = n(\br, t)- n(\br, 0)$, averaged over the $y$ direction and the ensemble, 
is shown for $\tI_1=0.21$ without the pump. (b) Dynamics in the presence of pump with $A_\md=0.8$. 
(c) Ensemble averaged phase difference $\phi$ at the junction for the cases in (a) and (b).
(d) The results of zeroth and first order expansion of Eq. \ref{eq:rsj} without the resistive term; see text.   
}
\label{Fig:mechanism}
\end{figure}

   Fig. \ref{Fig:dephasing}(a) shows the average total number of vortices in the system, 
determined by calculating the local phase winding around each lattice plaquette and averaging over both the time evolution and the initial ensemble,  
see End Matter for details.
The increase in vortex number coincides  with the slowdown and eventual reduction of the gain observed in Fig. \ref{Fig:gain}(a), 
confirming that vortex excitations are the primary dissipation mechanism limiting the amplification.  
During time evolution, the junction phase undergoes phase-slippage events, 
where the local phase difference at the barrier exceeds $\pi$, see Figs.  \ref{Fig:dephasing}(b, c). 
These events result in the creation of vortex-antivortex pairs near the junction, 
providing a dissipation channel for the injected pump energy. 
When averaged over many realizations, these stochastic slippage events give rise to dephasing in the imbalance dynamics, as shown in Fig.  \ref{Fig:dephasing}(c). 
This dephasing reflects the cumulative impact of vortex formation, which gradually suppresses coherent amplification and ultimately sets the limit for the achievable gain.

%
%
%

To understand the underlying nonlinear mechanism of the JPA we examine the time evolution of the density and the phase at the junction in Fig. \ref{Fig:mechanism}. 
Without pump, the barrier velocity modulation results in a smooth excitation of the density, which is created by the change of junction phase 
that is synchronized  with the signal frequency. 
When the pump field is on, there are higher-order density excitations that result from the modulation of the phase by the pump field. 
These can be explained with a perturbative solution of the equation $\tI_1 \cos(\omega_\ms t) = \bigl( 1+ A_\md \sin (\omega_\md t) \bigr) \sin \phi$, 
assuming a small modulation parameter $A_\md$ and expanding $\phi$ as $\phi= \phi_0 + A_\md \phi_1 + \mathcal{O}(A_\md^2) $. 
At zeroth order we obtain  $\phi_0 \approx \tI_1 \cos(\omega_\ms t) $, which explains the smooth oscillation of the phase without the pump. 
The first-order term gives $\phi_1 \approx - \tI_1 \cos(\omega_\ms t) \sin(\omega_\md t)\bigr)$, 
highlighting the intrinsic nonlinear mixing between the pump and signal field. 
In Fig. \ref{Fig:mechanism}(d), the perturbative solutions consistently capture the  dynamics of our atomic JPA.

{\it{Conclusions and outlook.}---}
We have studied the dynamics of a driven atomic Josephson junction (JJ) using classical-field simulations. 
The JJ is formed by coupling two two-dimensional atom clouds with a tunnel barrier, 
where we periodically modulate both the position (input signal) and the height (pump) of the barrier. 
By tuning the pump frequency to twice the Josephson plasma frequency,  
we have demonstrated parametric amplification for the signal. Our system works in the $\mHz$ frequency range, that is  beyond the operating range of the superconducting  JPAs, thereby offering potential for low-frequency signal detection and precision sensing of slowly varying fields, 
such as gravitational or magnetic fields \cite{Sheng2017,Javor2021,Somiya2023}. 
The current cold-atoms technology allows to accurately image the microscopic dynamics of the system \cite{Del_Pace2024, Bernhart2024}. 
In this case, we have an unprecedented access to the  density and phase {\it{microscopic}} dynamics at the junction responsible for the amplification and the intrinsic nonlinear mixing between the signal and pump fields. 
Our results can be experimentally probed using in-situ imaging and matter-wave interferometry, 
where the former enables measurement of the density imbalance and the latter allows detection of the underlying JPA mechanism. 
Our proposed protocol can be implemented in driven Josephson junctions experimentally realized in \cite{Del_Pace2024, Bernhart2024}.

Our findings have broad implications for the understanding of the basic mechanism on which parametric amplification relies in quantum systems. 
For the specific case of quantum electronics devices, while the microscopic  phase-density dynamics can be inferred through suitable Josephson current measurements, 
their direct experimental probing is out of the reach of current techniques, 
such as scanning probe microscopy.

{\it{Acknowledgments.}---}  
L. M. acknowledges funding by the Deutsche Forschungsgemeinschaft (DFG) in the excellence cluster  `Advanced Imaging of Matter’ - EXC 2056 - project ID 390715994 and by ERDF of the European
Union and `Fonds of the Hamburg Ministry of Science,
Research, Equalities and Districts (BWFGB)’.

\bibliography{References}

\clearpage

\appendix
\renewcommand{\thefigure}{A\arabic{figure}}
\renewcommand{\theequation}{A\arabic{equation}}
\setcounter{figure}{0}
\setcounter{equation}{0}

\section*{End Matter}

\begin{figure}
\includegraphics[width=1.0\linewidth]{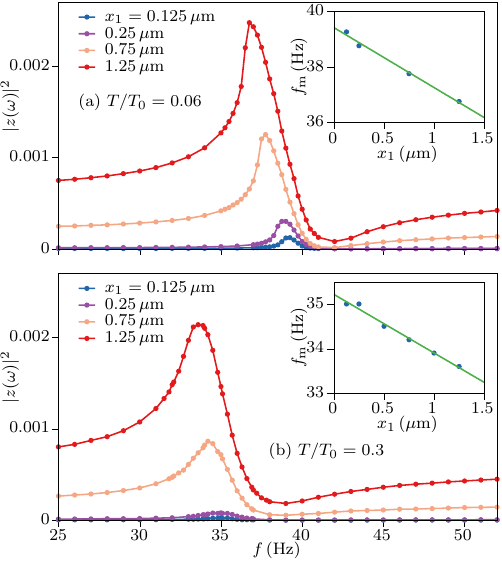}
\caption{(a, b) Power spectrum of the imbalance, $|z(\omega)|^2$, for different values of the modulation amplitude $x_1$, at $T/T_0= 0.06$ and $0.3$.
Inset shows the peak values of the spectrum as a function of $x_1$, 
where the linear fit (continuous line) in the limit of  $x_1 \rightarrow 0$ yields the resonance frequency 
$f_0=39.4\, \mHz$ and $35.2\, \mHz$ for $T/T_0= 0.06$ and $0.3$, respectively. 
}
\label{AppFig:freq}
\end{figure}

{\it{Josephson plasma frequency.}---} 
%
The protocol with periodic modulation of the barrier position results in coherent oscillations of the current across the junction, 
as demonstrated by the time evolution of the imbalance $z(t)$ across the junction in the main text.
Since the system behaves as a driven harmonic oscillator, a maximum response is expected at  the resonant frequency of the junction. 
To identify this resonance,  we calculate the power spectrum of the imbalance, $|z(\omega)|^2$, using a numerical Fourier transform. 
In Fig. \ref{AppFig:freq}, we show $|z(\omega)|^2$  determined for various values of the modulation amplitude $x_1$ 
at temperatures of $T/T_0= 0.06$ and $0.3$.  
The spectrum features a resonance peak, which we track at varying  $x_1$. 
Using a linear scaling approach, we determine the resonance frequency $f_0$ in the limit of  vanishing $x_1$, 
which gives $f_0= 39.4\, \mHz$ and $35.2\, \mHz$ for $T/T_0= 0.06$ and $0.3$, respectively.


%
\begin{figure}[t!]
\includegraphics[width=1.0\linewidth]{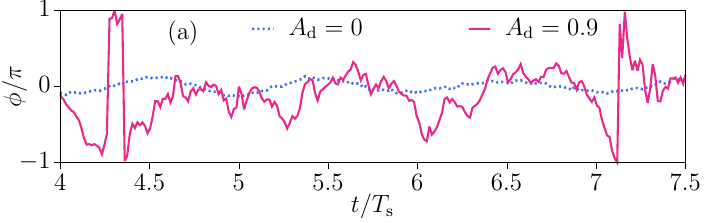}
\includegraphics[width=1.0\linewidth]{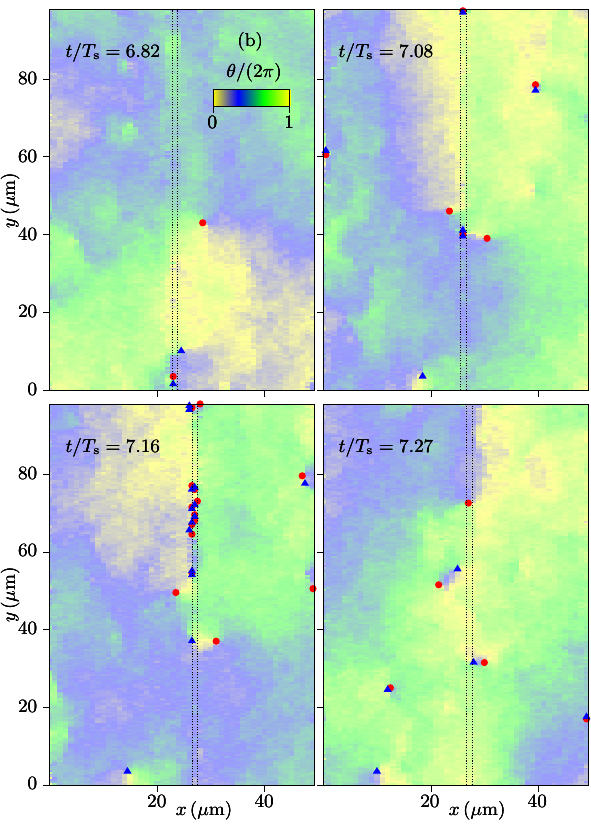}
\caption{
(a) Junction phase $\phi(t)$ for a single sample at $\tI_1=0.42$ with (continuous line) and without pump (dashed line), 
during the time interval between $4$ and $7.5$ cycles. $T_\ms$ is the signal oscillation period. 
(b) Snapshots of the phase distribution $\theta(x,y)$ at times near 7th cycle, for the same sample and parameters as in (a). 
The vertical dashed lines denote the location of the driven barrier. 
Vortices (triangles) and antivortices (dots) are identified based on the local phase winding; see text. 
}
\label{AppFig:dephasing}
\end{figure}
{\it{Dephasing dynamics.}---} 
Here, we elaborate on the dephasing dynamics observed in the density imbalance at large pump amplitudes $A_\md$. 
During time evolution, the junction phase exhibits strong fluctuations, 
which lead to the formation of vortex-antivortex pairs near the junction.
This is confirmed by analyzing the phase distribution in individual samples across the ensemble.
Fig. \ref{AppFig:dephasing} shows the time evolution of the phase field $\theta (x,y)$ during the interval in which the junction phase undergoes a strong change. 
Vortex excitations are identified  by computing the phase winding around the lattice plaquette of size $l\times l$ using $\sum_{\Box} \delta \theta(x,y) = \delta_x \theta(x,y) + \delta_y\theta(x+l,y)+\delta_x\theta(x+l,y+l)+\delta_y\theta(x,y+l)$, 
where $ \theta(x,y) $ is the phase field of $\psi(x,y)$ and  the phase differences between sites are taken to be $\delta_{x/y} \theta(x,y)  \in (-\pi, \pi]$. 
We identify a vortex (antivortex) by a phase winding of $2\pi$ ($-2\pi$). 
The total number of vortices $N_v$ is obtained by counting all vortices and antivortices in a given sample. 
This number is then averaged over the initial ensemble and the time evolution between $5$ and $15$ cycles.



\end{document}


\title{Supplemental Material for ``Atomic Josephson Parametric Amplifier''}
\author{Vijay Pal Singh}
\affiliation{Quantum Research Centre, Technology Innovation Institute, Abu Dhabi, UAE}
\author{Luigi Amico}
\affiliation{Quantum Research Centre, Technology Innovation Institute, Abu Dhabi, UAE}
\affiliation{Dipartimento di Fisica e Astronomia, Universit\`a di Catania, Via S. Sofia 64, 95123 Catania, Italy}
\affiliation{INFN-Sezione di Catania, Via S. Sofia 64, 95127 Catania, Italy}
\author{Ludwig Mathey}
\affiliation{Zentrum f\"ur Optische Quantentechnologien and Institut f\"ur  Quantenphysik, Universit\"at Hamburg, 22761 Hamburg, Germany}
\affiliation{The Hamburg Centre for Ultrafast Imaging, Luruper Chaussee 149, Hamburg 22761, Germany}
\date{\today}
\maketitle
%
%

%

%
\begin{figure}
\includegraphics[width=1.0\linewidth]{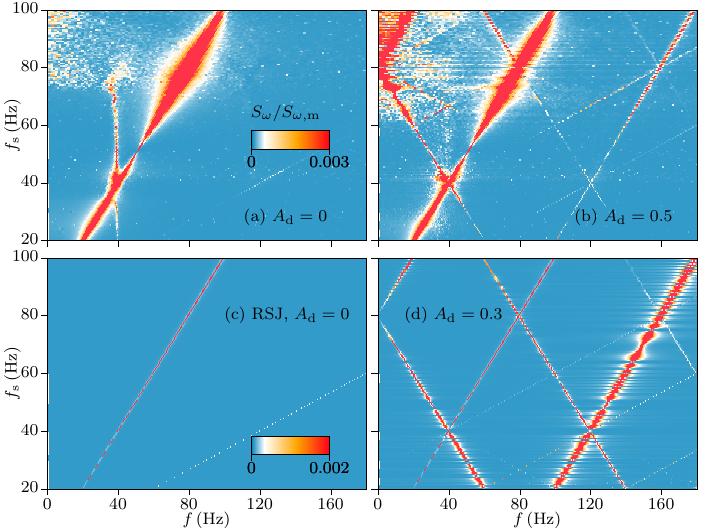}
\caption{Power spectrum of the imbalance,  $S_\omega=|z(\omega)|^2$, for signal frequencies between $f_\ms=20$ and $100\, \mHz$ at  $v_1/v_c=0.42$. 
For each column the spectrum is normalized by its maximum value  $S_{\omega, \mm}$. 
(a) Without the pump the spectrum mainly displays the central peak at $f_\ms=f$.  
(b) In the presence of pump there are peaks at $f_\md- f_\ms$ and $2f_\md- f_\ms$. 
(c, d) The results of the RSJ circuit model are shown for the same system parameters and $A_\md=0.3$; see text.
}
\label{Fig:spectrum}
\end{figure}

\section{Driven junction spectrum}
In Fig.  \ref{Fig:spectrum} we present the power spectrum of the imbalance,  $S_\omega=|z(\omega)|^2$, for signal frequencies between $f_\ms=20$ and $100\, \mHz$, using  $v_1/v_c=0.42$.  
The value of $v_1/v_c$ is chosen to be below $1$ to ensure a superfluid response, preventing  excitations associated with the resistive dynamics at the junction \cite{SinghShapiro}.    
Without the pump, the spectrum shows a dominant signal peak at $f_\ms=f$, with a resonant feature present at $f_\ms=f = 40 \, \mHz$. 
In the presence of the pump, the spectrum reveals two parametric resonances at  $f_\md- f_\ms$ and $2f_\md- f_\ms$. 
In the main text, we focus on the lowest resonance,  $f_\md- f_\ms$, as key operating mode of our amplifier.
To benchmark these results we compare them with the results of the driven RSJ model, 
which describes the junction dynamics and captures all key features of the power spectrum.

\begin{figure}
\includegraphics[width=1.0\linewidth]{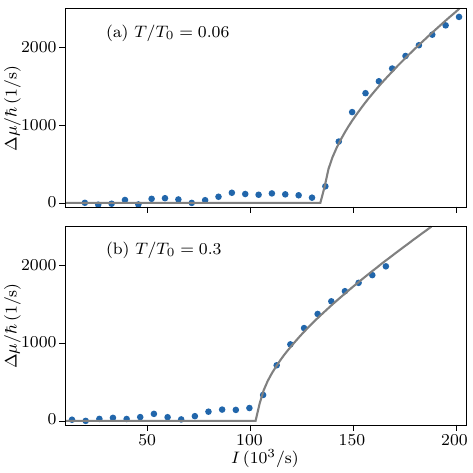}
\caption{
(a, b) Chemical-potential change $\Delta \mu$ as a function of the dc current $I$, using the final displacement of $\Delta x = 25\, \mum$, at $T/T_0=0.06$ and $0.3$.  
We fit the response with $\Delta \mu  = G^{-1} \sqrt{I^2 - I_c^2}$ (continuous line) to determine the
critical current $I_c$ and the conductance $G$. 
}
\label{Fig:iv}
\end{figure}

\begin{figure}
\includegraphics[width=1.0\linewidth]{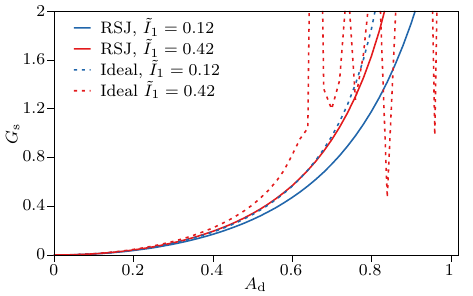}
\caption{Dimensionless signal gain $G_\ms$ as a function of the pump amplitude $A_\md$ 
for two different values of the signal current $\tilde{I}_1=0.12$ and $0.42$.  
Results from the RSJ model (continuous lines) are compared with those from the model without the conductance term (dashed lines),  referred to as the ideal circuit model. 
}
\label{Fig:gain}
\end{figure}

\section{Current-chemical potential characteristic}
To determine the current-chemical potential characteristic of the junction we follow the protocol provided in Ref. \cite{SinghShapiro}. 
We move the barrier at a constant velocity $v$ over a distance of $\Delta x = 25\, \mum$ and then calculate the final imbalance $z$ across the junction. 
This allows us to determine the velocity-imbalance characteristic, which we transform to the current-chemical potential characteristic using  $\Delta \mu =  N E_c z/2$ and $I(t) = (\bar{z} N/2) \times |v|/\Delta x$, where $\bar{z}$ is the equilibrium imbalance at the final location of the barrier. 
Fig. \ref{Fig:iv} shows the resulting current-chemical-potential characteristic of the junction for two different system temperatures of $T/T_0=0.06$ and $0.3$, with $T_0$ being an estimate of the critical temperature in dilute 2D Bose gases. 
We fit the response with the prediction of the RSJ circuit model, 
i.e., $\Delta \mu  = G^{-1} \sqrt{I^2 - I_c^2}$, with the critical current  $I_c$ and the conductance $G$ as fitting parameters \cite{SinghShapiro}. 
For $T/T_0=0.06$, we obtain  $I_c= 135\, (10^3/\ms)$ and $\hbar G= 59.8$, 
while for $T/T_0=0.3$,  we find  $I_c= 103\, (10^3/\ms)$ and $\hbar G= 62.8$.
Notably, the value of $I_c$ is reduced at higher temperature, 
which can be attributed to the suppression of bulk current by thermal fluctuations, as discussed in Ref. \cite{Singh2020jj}.
The onset of the dissipative regime above $I_c$ is associated with the creation of vortex-antivortex pairs \cite{SinghShapiro}. 
The motion of these excitations gives rise to a dissipative current,   
which acts as the primary mechanism of energy loss at the junction. This behavior can be regarded as an Ohmic-type dissipation, 
as $\Delta \mu$ increases approximately linearly with the current in the high-current regime  \cite{SinghShapiro}. 
The junction resistance is characterized by the conductance $G$, 
which is found to be similar at both temperatures.

\section{Driven RSJ circuit model}
As described in the main text, the driven RSJ circuit model reads
\begin{align}
 I_1 \cos(\omega_s t) = I_c\bigl( 1 + A_\md \sin(\omega_\md t) \bigr)  \sin \phi - G \Delta \mu,  \label{eq:rsj} 
\end{align}
where $\phi= \phi_L - \phi_R$ denotes the phase difference across the junction, and $G$ is the conductance.
The Josephson relation governing the phase dynamics is given by $\hbar \dot{\phi} = - \Delta \mu$, 
where $\Delta \mu$ plays the role of an effective voltage across the junction.
In this circuit model, the term $G \Delta \mu$ accounts for an Ohmic-type dissipation, 
which in our atomic junction arises from the creation of vortex-antivortex pairs, as discussed in the previous section. 
In Fig. \ref{Fig:gain}, we present results from the circuit model both with and without the conductance term. 
We refer to the latter as the ideal circuit model. 
We use the same parameters as in Fig. 4(a) of the main text. 
For  small pump amplitudes $A_\md$, the ideal and RSJ models yield consistent results.
However, at intermediate and high values of $A_\md$, the two models exhibit systematic deviations.  
In particular, the ideal model becomes unstable and shows diverging behavior at large signal currents, such as  $\tilde{I}_1 \equiv I_1/I_c= 0.42$.

\bibliography{References}